\documentclass[12pt]{article}
\usepackage{amsmath,amssymb,yfonts,esint}
\usepackage{wrapfig}
\usepackage{floatflt}
\usepackage{mathrsfs}

\usepackage[pdftex]{color,graphicx}

\usepackage{dsfont}
\newcommand{\be}{\begin{equation}}
\newcommand{\ba}{\begin{eqnarray}}
\newcommand{\ea}{\end{eqnarray}}
\newcommand{\nn}{\nonumber}

\def\e{\epsilon}

\def\G{\Gamma}

\def\ca{{\cal A}}
\def\cb{{\cal B}}

\def\ch{{\cal H}}

\newcommand{\pa}{\partial}

\fontfamily{yfrak}

\begin{document}

\vskip 15mm

\begin{center}

{\Large\bfseries Emergent Dirac Hamiltonians\\[2mm] in Quantum Gravity
%\\[2mm]
}

\vskip 4ex

Johannes \textsc{Aastrup}$\,^{a}$\footnote{email: \texttt{johannes.aastrup@uni-muenster.de}},
Jesper M\o ller \textsc{Grimstrup}\,$^{b}$\footnote{email: \texttt{grimstrup@nbi.dk}}\\ \& Mario \textsc{Paschke}$^{a}$\footnote{email: \texttt{mario.paschke@uni-muenster.de}}

%\& Ryszard \textsc{Nest}\,$^{c}$\footnote{email: \texttt{rnest@math.ku.dk}}

\vskip 3ex  

$^{a}\,$\textit{Mathematical Institute, University of M\"unster,\\ Einsteinstrasse 62, D-48149 M\"unster, Germany}
\\[3ex]
$^{b}\,$\textit{The Niels Bohr Institute, University of Copenhagen, \\Blegdamsvej 17, DK-2100 Copenhagen, Denmark}

%\\[3ex]
%$^{c}$ \textit{Mathematical Institute, University of Copenhagen,\\ Universitetsparken 5, DK-2100 Copenhagen, Denmark}
\end{center}

\vskip 3ex

\begin{abstract}

We modify the construction of the spectral triple over an algebra of holonomy loops by introducing additional parameters in form of families of matrices. These matrices generalize the already constructed Euler-Dirac type operator over a space of connections. We show that these families of matrices can naturally be interpreted as parameterizing foliations of 4-manifolds. The corresponding Euler-Dirac type operators then induce Dirac Hamiltonians associated to the corresponding foliation, in the previously constructed semi-classical states.

\end{abstract}

\newpage

%\tableofcontents

\section{Introduction}

It is still a major open task for any theory of quantum gravity to formulate an appropriate dynamical principle, i.e. the quantum analogue of  Einsteins equations, at the quantum level. No first principle in quantum theory is known which could lead to such a formulation. Such a principle should, in essence, require the independence of the construction of the theory under the choice of coordinate system and in particular the time coordinate. 

To mention only two issues which complicate the search for such a principle:
first, in many approaches to quantum gravity it is unclear how to formulate the semi-classical limit, thus obstructing a good physical intuition on the matter; second, the concrete dependence on the background is for some approaches highly unresolved.

In a recent publication \cite{Aastrup:2009et} we have constructed semi-classical states for a model of quantum gravity. Specifically, the model is given as a spectral triple construction over a configuration space of connections \cite{Aastrup:2005yk}-\cite{Aastrup:2009ux} , related to canonical quantum gravity formulated in terms of Ashtekar's variables \cite{Ashtekar:1986yd,Ashtekar:1987gu}. Essentially, the spectral triple is the formulation of an Euler-Dirac type operator {\it over} the infinite dimensional space of connections. The interaction between this operator and the algebra of holonomy loops, which is intrinsically noncommutative, reproduces the structure of the Poisson bracket of General Relativity and thus captures the kinematical part of quantum gravity. 
In \cite{Aastrup:2009et} we demonstrated that
 the construction contains certain normalizable semi-classical states on which the construction -- in particular the expectation value of the Euler-Dirac type operator -- descents to the Dirac Hamiltonian in 3+1 dimensions. Here, the lapse and shift fields which dictate a foliation of four-dimensional space-time, emerge from the semi-classical states.

However, a technical deficit concerning these semi-classical states did arise. Namely, the emergent inner product between spatial spinor fields depends on the lapse and shift fields. In this short note we show that this problem can be remedied by considering a more general class of Euler-Dirac type operators over connections. In particular, we permit a matrix action in the Euler-Dirac type operator which interacts directly with the noncommutativity of the holonomy loops. This action comes in the form of an infinite family of arbitrary two-by-two self adjoint matrices which can also be seen as a matrix valued field $M(x)$. These more general Euler-Dirac type operators $D_M$ lead, in the semi-classical analysis, again to the Dirac Hamiltonian in 3+1 dimensions, with the lapse and shift fields now emerging from this family of matrices.
 As we shall see, with this choice the normalization problem disappears.

Thus, we have now replaced the original model by a family of models, labelled by the fields $M(x)$ which, in the semi-classical approximation are identified as parameterizing the time-coordinate chosen for the model. As the subspace of physical states and the definition of observables thereon must not depend on the choice of a time-coordinate, this, to our point of view, clearly suggest a way to formulate a dynamical principle.

In fact, one constraint on physical states that seems to be required to this end,
$$
D_A D_{BC} = D_{AB}D_C \;,
$$
where $A,B,C$ are arbitrary matrix valued fields, resembles Einsteins equations in the semi-classical approximation. The exact analysis of this expression will be carried out in a future work.

Not that the transition from one foliation parametrized by $M(x)$ to another foliation $M'(x)$ is not unitarily implemented here, as had to be expected in quantum theory, in view of the Unruh effect.

Finally, this analysis clearly shows that the model should be understood in terms of a 3-dimensional theory living on a hyper-surface imbedded in a 4-dimensional manifold. This was previously not completely clear.

This note is intended as an addendum to the paper \cite{Aastrup:2009et}. Thus, we refer to \cite{Aastrup:2009et} for background information as well as notation.

\section{Semiclassical states}

We adopt the setup and notation from the paper \cite{Aastrup:2009et}. Thus, we consider a semi-finite spectral triple $(\cb,\ch,D)$ over a space $\ca$ of $SU(2)$ connections, based on an inductive system of 3-dimensional nested, cubic lattices, denoted $\{\G_n\}$. These graphs are embedded in a 3-dimensional oriented manifold $\Sigma$. The algebra $\cb$ is generated by based holonomy loops running in these graphs; the Hilbert space $\ch$ carries a representation of the algebra of holonomy loops together with an action of the Euler-Dirac type operator $D$. The Hilbert space $\ch$ is an inductive limit of intermediate Hilbert spaces $\ch_{\G_n}$ associated to graphs $\G_n$.  We denote by $\phi^t_{l_i}$ coherent states localized in the cotangent bundle of $SU(2)$ associated to edges $l_i$, and by $\phi^t_n$ products of coherent states over all copies of $SU(2)$ assigned to a given level $n$ in the inductive system of lattices. These coherent states are localized around a classical phase-space point given by a $SU(2)$ connection $A_j^a(x)$ and a densitized, inverse triad field $E_b^j(x)$. Also, let $\psi(x)$ be a two-by-two matrix-valued field.

\subsection{Semi-classical states for one vertex}

Consider three edges, denoted by $l_1$, $l_{2}$, $l_{3}$, all leading out of the same vertex, with three copies of $G$ associated to them, correspondingly. We assume that the edges belong to $\G_n \backslash \G_{n-1}$.
Consider the state in $\ch_{\G_n}$
\begin{eqnarray}
\Phi^t &=&\Big(\frac{ \mathrm{i} }{5} {\bf e}_1^c   {\bf e}_2^a  {\bf e}_3^b (    \delta^{ab} \sigma^c +   \delta^{ac} \sigma^b  + \delta^{bc} \sigma^a)  
%-    {\bf e}_1^a   {\bf e}_2^a   + {\bf e}_1^a  {\bf e}_3^a- {\bf e}_2^a  {\bf e}_3^a
   \psi(v_0)
\nn\\&&
+  {\bf e}_1^a   {\bf e}_2^a g_3 \psi(v_3)   - {\bf e}_1^a  {\bf e}_3^a g_2 \psi(v_2)
%\ nn\\&&
+ {\bf e}_2^a  {\bf e}_3^a  g_1 \psi(v_1)  \Big)   \phi^t_{l_1}\phi^t_{l_2}\phi^t_{l_3}\;.
\label{darwin}
\end{eqnarray}
where the enumeration of the vertices are as show in figure 1 and where $ {\bf e}_j^a$ are elements of the Clifford algebra associated to the $j$'th copy of $SU(2)$. Expression (\ref{darwin}) is the state also analyzed in \cite{Aastrup:2009et}.

\begin{figure}[t]
\begin{center}
\resizebox{!}{4cm}{
 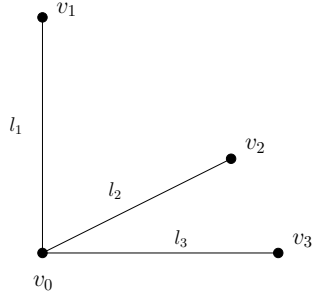}
\end{center}
\caption{\it Three edges connected in the vertex $v_0$.}
\end{figure}
%Now comes the novel ingredient particular to this paper. 
The original Euler-Dirac type operator $D$ does not act on the matrix factor in $\ch$. However, such an action can easily be introduced by altering the operator to
\begin{equation}
D_M = \sum_{j,a} {\bf e}_j^a d_{{\bf e}_j^a}  M(j)
\label{DDDMMM}
\end{equation}
where $M(j)$ are arbitrary two-by-two self-adjoint matrices. Here, $d_{{\bf e}_j^a}$ is the left invariant vector field on the copy of $SU(2)$ associated to the edge $l_j$ corresponding to the generator in $\mathfrak{su}(2)$ with index $a$. The original Euler-Dirac type operator is obtained by setting $M(j)\equiv  \mathds{1}_2$ for all $j$.
If we calculate the expectation value of $D_M$ over the states (\ref{darwin}) we find
\begin{eqnarray}
\langle \Phi^t \vert D_M \vert \Phi^t \rangle =  a_n 2^{-3n} \psi(\sigma^a E^a_i  M(i)\nabla_i  +\nabla_i  M(i)\sigma^a E^a_i )\psi
\label{ddd}
\end{eqnarray}
where $\nabla_i = \pa_i + A_i $, $i\in\{1,2,3\}$ and where we left out zero order terms. This expression only holds when $n$ is taken to infinity to form the derivative and it depends on a choice of the scaling parameters, 
$
a_n = 2^{3n}
$,
see \cite{Aastrup:2009et}.
Equation  (\ref{ddd}) is, up to zero order terms, the expression which in \cite{Aastrup:2009et} lead to the Dirac Hamiltonian in 3+1 dimensions. The lapse and shift fields come from the matrices $M(i)$ through the expansion
$
M(i) = N(i) \mathds{1} +\mathrm{i} N^a(i) \sigma^a\;,
$
where $N(i)$ and $N^a(i)$ are interpreted as smooth fields on $\Sigma$ taking values at $v_i$. Referring to the numbering in figure 1, we write $M(1)=M(2)=M(3)\equiv M(v_0)$. Thus, $N(v_i)$ will be the lapse field and $N^m(v_i)= e^m_a N^a(v_i)$ will be the shift field.

\subsection{Semi-classical states over $\overline{\ca}$}

We follow again \cite{Aastrup:2009et}. 
At the $n$'th level in the inductive system of lattices we write down the state in $\ch_{\G_n}$
\begin{equation}
\Phi^t_n(\ca_{\G_n}) =  2^{-3n/2}\left(\sum_{v_j} \Psi_{v_j} \right)\phi_n^t \;,
\label{leveln}
\end{equation}
with
\begin{eqnarray}
\Psi_{v_j} &=&
\frac{ \mathrm{i} }{5} {\bf e}_1^c   {\bf e}_2^a  {\bf e}_3^b (    \delta^{ab} \sigma^c +   \delta^{ac} \sigma^b  + \delta^{bc} \sigma^a)  
%-    {\bf e}_1^a   {\bf e}_2^a   + {\bf e}_1^a  {\bf e}_3^a- {\bf e}_2^a  {\bf e}_3^a
 \psi(v_{j})
\nn\\&&
+  {\bf e}_1^a   {\bf e}_2^a g_3 \psi(v_{j_3})   - {\bf e}_1^a  {\bf e}_3^a g_2 \psi(v_{j_2})
%\nn\\&&
+ {\bf e}_2^a  {\bf e}_3^a  g_1 \psi(v_{j_1})
\label{labellabel}
\nonumber
\end{eqnarray}
where $\{v_j,v_{j_1}, v_{j_2} ,v_{j_3}\}$ denote vertices corresponding to three edges coinciding at the vertex $v_j$. The sum in (\ref{leveln}) runs over a certain set of vertices in $\G_n$ which are new at the $n$'th level, see \cite{Aastrup:2009et}.

With (\ref{leveln}) we have a sequence $\{\Phi^t_n\}$ of states in $\ch$ and we can calculate the limit of the expectation value of $D_M$ on these states. We find
\begin{eqnarray}
\lim_{n\rightarrow\infty} \lim_{t\rightarrow 0}  \langle \bar{\Phi}^t_n \vert  D_M \vert \Phi^t_n  \rangle 
\nn\\
&&\hspace{-40mm}=  \int_\Sigma d^3x\bar{ \psi}(x) \left( \frac{1}{2}(\sqrt{g}N \sigma^a e_a^m\nabla_m + N\nabla_m \sqrt{g}\sigma^a e^m_a) +\mathrm{i}\sqrt{g} N^m\pa_m \right) \psi(x) 
\nonumber\\
&& \hspace{-30mm}+\;\; \mbox{\it zero order terms}
   \;,
\label{dex2}
\end{eqnarray}
which, up to zero order terms, equals the Dirac Hamiltonian in 3+1 dimensions and thus matches results in  \cite{Aastrup:2009et}. Therefore, the discussion and remarks written in \cite{Aastrup:2009et} equally applies here.
The central difference to the result in \cite{Aastrup:2009et} is that the states no longer depend on the lapse and shift fields. Thus, the inner product 
$$
\lim_{n\rightarrow\infty} \lim_{t\rightarrow 0}  \langle \bar{\Phi}^t_n  \vert \Phi^t_n  \rangle 
=  \int_\Sigma d^3x\bar{ \psi}(x) \psi(x)\;,
$$
is independent of the lapse and shift fields, in contrast to the construction in \cite{Aastrup:2009et}.

Let us here add that the spatial Dirac operator appearing in (\ref{dex2}) (and also found in \cite{Aastrup:2009et}) is, when interpreted in terms of Ashtekars variables, not an ordinary 3-dimensional Dirac operator. Rather, it involves the Ashtekar connection and thereby the extrinsic curvature of $\Sigma$ in the space-time manifold $M$. Thus, it resembles a 4-dimensional Dirac operator restricted to a 3-dimensional hyper-surface.

\section{Discussion}

The way the lapse and shift fields emerge from the general Euler-Dirac type operator $D_M$ seems to point towards a natural equivalence principle: to incorporate invariance under the choice of foliation in the construction invariance under the transformation
\[
\{M(i)\}\rightarrow \{M'(i)\}\;.
\]
should be implemented.

Let us consider for a moment the constraints of General Relativity.
The diffeomorphism and Hamilton constraints have the form (we omit a factor $\sqrt{g}^{-1}$)
\[
N^a E_a^m E_b^n F^b_{mn}\;,\quad N \e^{abc} E^m_a E^n_b F^c_{mn}\;,
\]
where $F$ is the field strength tensor of the Ashtekar connection. If we add them we obtain an expression
%$
%E^m_a E^n_b F^c_{mn}( N \e^{abc} + N^a \d^{bc})\;,
%$
which can be seen as coming from the trace of 
\begin{equation}
%E^m_a \sigma^a E^n_b \sigma^b F_{mn}^c \sigma^c (N + \mathrm{i}N^d \sigma^d) = 
E^m_a (\mathrm{i}\sigma^a) E^n_b (\mathrm{i}\sigma^b) F_{mn}^c (\mathrm{i}\sigma^c) M\;.
\label{mmmm}
\end{equation}
Let us, at a heuristic level, see what operator this quantity might correspond to. First, we "backtrack" and write ${\rm{i}}d_{{\bf e}_a^m}$ instead of $E_a^m$ (using the Poisson structure etc., see \cite{Aastrup:2009et}). Second, we exchange two $\sigma$'s with Clifford elements. This may be implemented through factors ${\rm{i}}\sigma^a {\bf e}_a^m$ in a semi-classical state, like the ones in (\ref{labellabel}). With these formal operations expression (\ref{mmmm}) looks like
\[
(\rm{i}{\bf e}^m_a d_{{\bf e}^m_a} )(\rm{i} {\bf e}^n_b d_{{\bf e}^n_b} ) F_{mn}^c (\mathrm{i}\sigma^c) M
\label{MMM}
\]
Also, the $F$ could be seen as coming from the state through factors of $g_i$, referring again to the state (\ref{labellabel}). This crude analysis suggest that we are looking for an expression quadratic in the Euler-Dirac type operator $D$. This, however, cannot be true since (\ref{mmmm}) involves cross-terms with left-invariant vector fields corresponding to different edges. Such terms cannot\footnote{The reason for this being the choice of the Clifford algebra for the construction of $D$.} come from the square of $D$. If, however, we consider instead the more general Euler-Dirac type operators $D_M$ then cross-terms do in fact appear. Writing
\[
D_M D_M = \sum_i {\bf e}_i^a {\bf e}_i^b d_{{\bf e}_i^a}d_{{\bf e}_i^b} M(i)M(i) + \sum_{i \not= j} {\bf e}_i^a {\bf e}_j^b d_{{\bf e}_i^a}d_{{\bf e}_j^b} [M(i),M(j)]
\]
we see that unless all the matrices $M(i)$ are diagonal in the same basis of $\mathbb{C}^2$, cross terms which involve vector fields associated to different edges will appear.
Further, to avoid the first diagonal terms, we can write down the expression
\[
D_A D_{BC} = D_{AB} D_C
\]
where $A,B,C$ are arbitrary matrix valued fields. This expression involves only cross terms and will resemble, in an appropriate semi-classical limit, the Einstein equations.
This expression will be analyzed in a future work.

Note that the operator $D_M$ does, strictly speaking, not satisfy the requirements of a spectral triple since its commutator with the algebra of holonomy loops is no longer bounded.

Finally, we should point out that the foliation does not only depend on the choice of the field $M(x)$ but also on the semi-classical state through the (inverse) triad field $e_a^m(x)$.\\

\noindent{\bf\large Acknowledgements}\\

\noindent
J.A. and M.P. were supported by the SFB 478 grant  "Geometrische Strukturen in der Mathematik" of the Deutsche Forschungsgemeinschaft. We would like to thank Ryszard Nest for discussions.

\end{document}